\documentclass[floatfix, preprint, onecolumn, prd,
showpacs, showkeys, preprintnumbers, nofootinbib, superscriptaddress]{revtex4-1}
\usepackage{natbib}
\usepackage{times}
\usepackage{amssymb,amsbsy,amsmath,amsfonts}
\usepackage{graphicx}
\usepackage{float}
\usepackage{color}
\usepackage{morefloats}
\usepackage{rotating}
\usepackage{srcltx}
\usepackage{slashed}
\usepackage{subfigure}
\usepackage{multirow}
\usepackage{verbatim}
\usepackage{hyperref}
\usepackage{tabularx}

\graphicspath{{./}{./img/}{./fig/}{./image/}{./figure/}{./picture/}}

\begin{document}

\title{Virtual decuplet effects on octet baryon masses in covariant baryon chiral perturbation theory}

\author{Xiu-Lei Ren}
\affiliation{School of Physics and
Nuclear Energy Engineering, Beihang University, Beijing 100191, China}

\author{Lisheng Geng}
\email[E-mail me at: ]{lisheng.geng@buaa.edu.cn}
\affiliation{School of Physics and
Nuclear Energy Engineering, Beihang University, Beijing 100191, China}

\author{Jie Meng}
\affiliation{School of Physics and
Nuclear Energy Engineering, Beihang University, Beijing 100191, China}
\affiliation{State Key Laboratory of Nuclear Physics and Technology, School of Physics, Peking University, Beijing 100871, China}
\affiliation{Department of Physics, University of Stellenbosch, Stellenbosch, South Africa}

\author{H. Toki}
\affiliation{Research Center for Nuclear Physics (RCNP), Osaka University, Ibaraki, Osaka 567-0047, Japan}

\begin{abstract}
  We extend a previous analysis of the lowest-lying octet baryon masses in covariant baryon chiral perturbation theory (ChPT) by explicitly taking into account the contribution of the virtual decuplet baryons. Up to next-to-next-to-next-to-leading order (N$^3$LO), the effects of these heavier degrees of freedom are systematically studied.
  Their effects on the light-quark mass dependence of the octet baryon masses
  are shown to be relatively small and can be absorbed by the available low-energy constants up to N$^3$LO.
  Nevertheless, a better description of the finite-volume corrections of the lattice QCD data can be achieved,
   particularly those with small $M_\phi L$ ($<4$), which is demonstrated by a careful study of  the NPLQCD and QCDSF-UKQCD small-volume data.  Finally, we show that the predicted pion- and strangeness-baryon sigma terms are only slightly changed by the inclusion of the virtual decuplet baryons.

\end{abstract}


\date{\today}

\maketitle

\section{\label{sec:intro}Introduction}

In recent years, lattice chromodynamics (LQCD) simulations~\cite{Wilson:1974sk,Gattringer2010} have made remarkable progress in studies of non-perturbative strong-interaction physics (e.g., see Ref.~\cite{Colangelo:2010et}). Accurate computations of the lowest-lying octet baryon spectrum with $n_f=2+1$ dynamical simulations have been reported by several LQCD collaborations~\cite{Durr:2008zz,Alexandrou:2009qu,Aoki:2008sm,Aoki:2009ix,WalkerLoud:2008bp,Lin:2008pr, Bietenholz:2010jr,Bietenholz:2011qq,Beane:2011pc}. Nonetheless, most LQCD calculations employed larger than physical light-quark masses~\footnote{Up to now, only a few lattice collaborations have performed simulations at or close to the physical light-quark masses with fixed $m_s\simeq m_s^{\rm phys.}$~\cite{Durr:2010aw,Bazavov:2012uw}.}  and finite volume, therefore the obtained results have to be extrapolated to the physical point by performing the so-called ``chiral extrapolation"~\cite{Leinweber:2003dg,Bernard:2003rp,Procura:2003ig,Bernard:2005fy} and taking into account the finite-volume corrections (FVCs)~\cite{Gasser:1986vb,Gasser:1987zq}.

Chiral perturbative theory, as the low energy effective field theory of QCD~\cite{Weinberg:1978kz,Gasser:1983yg,Gasser:1984gg,Gasser:1987rb,Leutwyler:1994fi, Bernard:1995dp,Pich:1995bw,Ecker:1994gg,Pich:1998xt,Bernard:2006gx,Bernard:2007zu,Scherer2012b}, provides an appropriate framework to study the light-quark mass dependence and the lattice volume dependence of LQCD results. In the last decades, baryon chiral perturbation theory (BChPT)
has been applied to study the ground state (g.s.) octet baryon masses with~\cite{Jenkins:1991ts,Bernard:1993nj,WalkerLoud:2004hf,Lehnhart:2004vi} and without~\cite{Borasoy:1996bx,Ellis:1999jt,Frink:2004ic,Frink:2005ru} the explicit inclusion of the intermediate decuplet resonances.
Recently, the $n_f=2+1$ LQCD simulations of the lowest-lying baryon masses at different combinations of light-quark and strange-quark masses have been studied with
 the SU(3) BChPT up to $\mathcal{O}(p^4)$~\cite{Ishikawa:2009vc,Young:2009zb,MartinCamalich:2010fp,Bruns:2012eh,Semke:2011ez,
Semke:2012gs,Lutz:2012mq,Ren:2012aj}. It is found that the non-relativistic heavy baryon (HB) ChPT converges rather slowly up to next-to-next-to-leading order (NNLO) ~\cite{WalkerLoud:2008bp,Ishikawa:2009vc}.
The situation is much better in the finite range regulator (FRR) method ~\cite{Young:2009zb} and
in the extended-on-mass-shell (EOMS) BChPT~\cite{MartinCamalich:2010fp}. Up to N$^3$LO,
the infrared BChPT~\cite{Bruns:2012eh}, the partial summation scheme~\cite{Semke:2011ez,Semke:2012gs,Lutz:2012mq}
, and the EOMS BChPT~\cite{Ren:2012aj} all seem to be able to describe the LQCD data, but only in Ref.~\cite{Ren:2012aj}  finite-volume effects were taken into account self-consistently and all the relevant LECs were determined from the LQCD and experimental data.

Using the covariant BChPT with the EOMS scheme~\cite{Gegelia:1999gf,Fuchs:2003qc}, we have performed a series of studies on the LQCD mass results~\cite{MartinCamalich:2010fp,Geng:2011wq,Ren:2012aj}. In Ref.~\cite{MartinCamalich:2010fp}, it is shown that the EOMS BChPT can provide a better description of the PACS-CS~\cite{Aoki:2008sm} and LHPC~\cite{WalkerLoud:2008bp} data and is more suitable for chiral extrapolation than the HBChPT up to NNLO. In Ref.~\cite{Geng:2011wq}, the NNLO EOMS BChPT is used to analyze the NPLQCD~\cite{Beane:2011pc} lattice volume dependent results at fixed quark masses. It is shown that for moderate $M_\pi L$, FVCs need to be taken into account in order to describe the lattice data and they
are also helpful to constrain the value of some relevant LECs.
In Ref.~\cite{Ren:2012aj}, we performed a simultaneous fit of all the publicly available LQCD data from the PACS-CS~\cite{Aoki:2008sm}, LHPC~\cite{WalkerLoud:2008bp}, HSC~\cite{Lin:2008pr}, QCDSF-UKQCD~\cite{Bietenholz:2011qq} and NPLQCD~\cite{Beane:2011pc} collaborations using the N$^3$LO EOMS BChPT with the FVCs taken into account self-consistently. It is shown that the covariant SU(3) BChPT converges as expected with clear improvement order by order, and all the lattice simulations are consistent with each other, although their setups are quite different. It should be mentioned that the contributions of the virtual decuplet baryons were not explicitly included in the N$^3$LO BChPT calculation of Ref.~\cite{Ren:2012aj} with the assumption that the decuplet effects can not be disentangled from those of the relevant LECs.

However, in the SU(3) BChPT one should be careful about the contributions of the decuplet resonances since the average mass gap between the baryon octet and the baryon decuplet $\delta=m_D-m_0\sim 0.3$ GeV is similar to the pion mass and  well bellow those of the kaon and eta mesons, $M_K, M_{\eta}$. In Ref.~\cite{Jenkins:1991ts}, HBChPT was enlarged to include the decuplet and applied to calculate the octet baryon masses  up to $\mathcal{O}(p^3)$, and the importance of the spin-$3/2$ fields was pointed out. However, it was shown in Ref.~\cite{Bernard:1993nj} that the effects of the virtual decuplet on the octet baryon masses actually start out at $\mathcal{O}(p^4)$ in the same framework. For the spin-dependent quantities, the virtual decuplet contributions are found to be important in HBChPT, such as  the magnetic moments~\cite{Puglia:2000jy} and the axial vector form factors~\cite{Jenkins:1991es,Luty:1993gi}.  In the EOMS BChPT,  the effects of the virtual decuplet are found to be negligible  for the magnetic moments of the octet baryons if the ``consistent'' coupling scheme for the octet-decuplet-pseudoscalar coupling is adopted~\cite{Geng:2009hh}. On the hand, up to NNLO, the virtual decuplet contributions seem to play an important role in describing the NPLQCD  volume-dependent data~\cite{Geng:2011wq}  and in
the determination of the baryon sigma terms~\cite{Alarcon:2012nr}. Therefore, it is necessary to study  the effects of the virtual decuplet on the
light-quark mass and volume dependence of the LQCD data and on the determination of the baryon sigma terms at N$^3$LO.

In this work, we explicitly take into account the contributions of the virtual decuplet resonances to the g.s. octet baryon masses in the EOMS BChPT up to N$^3$LO. The finite-volume corrections from the virtual decuplet are calculated self-consistently. Through a simultaneous fit of the publicly available $n_f=2+1$ LQCD data from the PACS-CS~\cite{Aoki:2008sm}, LHPC~\cite{WalkerLoud:2008bp}, HSC~\cite{Lin:2008pr}, QCDSF-UKQCD~\cite{Bietenholz:2011qq} and NPLQCD~\cite{Beane:2011pc} collaborations, we perform a systematic study of the intermediate decuplet effects on the chiral extrapolation of and the FVCs to the octet baryon masses, and on the determination of the octet baryon sigma terms
through the Feynman-Hellmann theorem.

This paper is organized as follows. In Sec.~\ref{SecII}, we collect the chiral effective Lagrangians involving the decuplet baryons and calculate their contributions to the octet baryon masses in the EOMS BChPT up to N$^3$LO. In Sec.~\ref{SecIII}, we perform a simultaneous fit of the LQCD data and study the effects of the virtual decuplet baryons in detail.  A short summary is given in Sec.~\ref{SecIV}.

\section{Theoretical Framework}\label{SecII}
\subsection{Chiral effective Lagrangians involving the decuplet baryons}

The baryon decuplet consists of a SU(3)-flavor multiplet of spin-$3/2$ resonances, which are  represented with the {\it Rarita-Schwinger} field~$T^{abc}\equiv T_{\mu}^{abc}$ (each element of $T_{\mu}^{abc}$ is a four-component Dirac spinor). The physical fields are assigned to the tensor as $T^{111}=\Delta^{++}$, $T^{112}=\Delta^+/\sqrt{3}$, $T^{122}=\Delta^0/\sqrt{3}$, $T^{222}=\Delta^-$,
$T^{113}=\Sigma^{*+}/\sqrt{3}$, $T^{123}=\Sigma^{*0}/\sqrt{6}$, $T^{223}=\Sigma^{*-}/\sqrt{3}$, $T^{133}=\Xi^{*0}/\sqrt{3}$, $T^{233}=\Xi^{*-}/\sqrt{3}$, and $T^{333}=\Omega^-$.

The covariant free Lagrangian for the decuplet baryons is
\begin{equation}
  \mathcal{L}_T=\bar{T}_{\mu}^{abc}\left(i\gamma^{\mu\nu\alpha}D_{\alpha} - m_D\gamma^{\mu\nu}\right)T_{\nu}^{abc},
\end{equation}
where $m_D$ is the decuplet-baryon mass in the chiral limit and $D_{\nu}T_{\mu}^{abc}=\partial_{\nu}T_{\mu}^{abc} + (\Gamma_{\nu}, T_{\mu})^{abc}$, $\Gamma_{\nu}$ being the chiral connection (see, e.g., Ref.~\cite{Scherer2012b}) and with the definition $(X, T_{\mu})^{abc}\equiv (X)^a_dT_{\mu}^{dbc} + (X)^b_dT_{\mu}^{adc} + (X)^c_dT_{\mu}^{abd}$. In the last and following Lagrangians, we always apply the Einstein notation to sum over any repeated SU(3)-index denoted by latin characters $a$, $b$, $c$, $\cdots$, and $(X)_b^a$ denotes the element of row $a$ and column $b$ of the matrix representation of $X$.
The totally antisymmetric gamma matrix products are defined as: $\gamma^{\mu\nu}=\frac{1}{2}\left[\gamma^{\mu}, \gamma^{\nu}\right]$, $\gamma^{\mu\nu\alpha}=\frac{1}{2}\left\{\gamma^{\mu\nu},\gamma^{\alpha}\right\}= -i\varepsilon^{\mu\nu\alpha\beta}\gamma_{\beta}\gamma_5$, with the following conventions: $g^{\mu\nu}={\rm diag}(1,-1,-1,-1)$, $\varepsilon_{0,1,2,3}=-\varepsilon^{0,1,2,3}=1$ and $\gamma_5=i\gamma_0\gamma_1\gamma_2\gamma_3$.

The $\mathcal{O}(p^2)$ chiral Lagrangian for the decuplet baryons is:
\begin{equation}
  \mathcal{L}_T^{(2)}=\frac{t_0}{2}\bar{T}_{\mu}^{abc}g^{\mu\nu}T_{\nu}^{abc}\langle\chi_{+}\rangle + \frac{t_D}{2}\bar{T}_{\mu}^{abc}g^{\mu\nu}(\chi_+, T_{\nu})^{abc},
\end{equation}
with $\chi_+=2\chi=4B_0{\rm diag}(m_l,~m_l,~m_s)$ introducing the explicit chiral symmetry breaking, where
 $m_l$ and $m_s$ are the average light-quark and strange-quark masses. The parameters $t_0$, $t_D$ are two unknown LECs.

Up to $\mathcal{O}(p^3)$ the chiral effective Lagrangian, describing the interaction of the octet and decuplet baryons with the pseudoscalar mesons, can be written as~\cite{Geng:2009hh}
\begin{equation}\label{eq:LBDP1}
  \mathcal{L}_{\phi BT}^{(1)} = \frac{i\mathcal{C}}{m_DF_{\phi}}\varepsilon^{abc}(\partial_{\alpha}\bar{T}_{\mu}^{ade}) \gamma^{\alpha\mu\nu}B_c^e\partial_{\nu}\phi_b^d + {\rm H.c.},
\end{equation}
where we have used the so-called ``consistent'' coupling scheme for the octet-decuplet-pseudoscalar vertices~\cite{Pascalutsa:1998pw,Pascalutsa:1999zz}. The $\phi$ and $B$ are the SU(3) matrix representations of the pseudoscalar mesons and of the octet baryons. The coefficient $F_{\phi}$ is the meson-decay constant in the chiral limit, and $\mathcal{C}$ denotes the $\phi BT$ coupling. 

The propagator of the spin-$3/2$ fields in $d$ dimensions has the following form~\cite{Pascalutsa:2005nd}
\begin{equation}
  S^{\mu\nu}(p) = -\frac{\slashed p+ m_D}{p^2-m_D^2+i\epsilon} \left[g^{\mu\nu}-\frac{1}{d-1}\gamma^{\mu}\gamma^{\nu} - \frac{1}{(d-1)m_D}\left(\gamma^{\mu}p^{\nu}-\gamma^{\nu}p^{\mu}\right) - \frac{d-2}{(d-1)m_D^2}p^{\mu}p^{\nu}\right].
\end{equation}

\subsection{Virtual decuplet contributions to the octet baryon masses}

\begin{figure}[t]
  \centering
  \includegraphics[width=15cm]{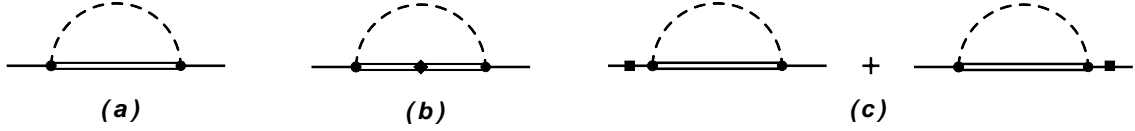}\\
  \caption{Feynman diagrams contributing to the octet baryon masses with the intermediate decuplet resonances. The solid lines correspond to octet baryons, the double lines to decuplet baryons, and the dashed lines denote pseudoscalar mesons. Black dots indicate an insertion from the dimension one chiral Lagrangian (Eq.~(\ref{eq:LBDP1})), and black boxes (diamonds) indicate $\mathcal{O}(p^2)$ mass insertions.}\label{Fig1:FeynmanD}
\end{figure}

 Because the baryon mass, which is of the same order as the chiral symmetry breaking scale $\Lambda_{\rm ChPT}$, does not vanish in the chiral limit, a systematic power-counting (PC) is destroyed beyond the leading order calculation in BChPT~\cite{Gasser:1987rb}. In order to restore the chiral power-counting, the extended-on-mass-shell (EOMS) renormalization scheme was proposed~\cite{Gegelia:1999gf,Fuchs:2003qc}.  The essence of the EOMS scheme is to perform an additional subtraction of power-counting breaking (PCB) pieces beyond the $\widetilde{\rm MS}$ or $\overline{\rm MS}$ renormalization scheme. Different from the infrared (IR) BChPT~\cite{Becher:1999he} and HBChPT~\cite{Jenkins:1990jv}, the EOMS BChPT is not only covariant, but also satisfies all analyticity and symmetry constraints (see, e.g., Ref.~\cite{Geng:2013xn} ). In addition, it converges relatively faster~\cite{Geng:2008mf,Geng:2009hh,Geng:2009ys}. In this work we use the EOMS scheme to remove the PCB terms from the one-loop diagrams.

\begin{table}[t]
  \centering
  \caption{Coefficients of the NNLO virtual decuplet contribution to the self-energy of the octet baryons (Eq.~(\ref{masseqD})).}\label{Coef:nlo}
  \begin{tabular}{lllll}
     \hline\hline
       & $N$ & $\Lambda$ & $\Sigma$ & $\Xi$ \\
     \hline
     $\xi_{B,\pi}^{(a)}$ & $\frac{16}{3}\mathcal{C}^2$ & $4\mathcal{C}^2$ & $\frac{8}{9}\mathcal{C}^2$ & $\frac{4}{3}\mathcal{C}^2$\\
     $\xi_{B,K}^{(a)}$ & $\frac{4}{3}\mathcal{C}^2$ & $\frac{8}{3}\mathcal{C}^2$ & $\frac{40}{9}\mathcal{C}^2$ & $4\mathcal{C}^2$ \\
     $\xi_{B,\eta}^{(a)}$ & $0$ & $0$ & $\frac{4}{3}\mathcal{C}^2$ & $\frac{4}{3}\mathcal{C}^2$ \\
     \hline\hline
   \end{tabular}
\end{table}

\begin{table}[h!]
  \centering
  \caption{Coefficients of loop diagrams (Fig.~\ref{Fig1:FeynmanD}(b/c)) to the self-energy of the octet baryons~(Eq.~(\ref{masseqD})).}\label{Coef:nnlo}
  \begin{tabular}{llll}
     \hline\hline
       $N$ & $\Lambda$ & $\Sigma$ & $\Xi$ \\
     \hline
     $\xi_{N\Delta,\pi}^{(b/c)}=4\mathcal{C}^2$ & $\xi_{\Lambda\Sigma^*,\pi}^{(b/c)}=3\mathcal{C}^2$ & $\xi_{\Sigma\Delta,K}^{(b/c)}=\frac{8}{3}\mathcal{C}^2$ & $\xi_{\Xi\Sigma^*,K}^{(b/c)}=\mathcal{C}^2$ \\
     $\xi_{N\Sigma^*,K}^{(b/c)}=\mathcal{C}^2$ & $\xi_{\Lambda\Xi^*,K}^{(b/c)}=2\mathcal{C}^2$ & $\xi_{\Sigma\Sigma^*,\pi}^{(b/c)}=\frac{2}{3}\mathcal{C}^2$ & $\xi_{\Xi\Xi^*,\pi}^{(b/c)}=\mathcal{C}^2$ \\
       &   & $\xi_{\Sigma\Sigma^*,\eta}^{(b/c)}=\mathcal{C}^2$ & $\xi_{\Xi\Xi^*,\eta}^{(b/c)}=\mathcal{C}^2$ \\
       &   & $\xi_{\Sigma\Xi^*,K}^{(b/c)}=\frac{2}{3}\mathcal{C}^2$ & $\xi_{\Xi\Delta^-,K}^{(b/c)}=2\mathcal{C}^2$\\
     \hline\hline
   \end{tabular}
\end{table}

\begin{table}[h!]
  \centering
  \caption{Coefficients of LO contribution to the self-energy of the decuplet baryons (Eq. (\ref{mass:D2})).}
  \label{Coef:dlo}
  \begin{tabular}{ccccc}
    \hline\hline
     & $\Delta$ & $\Sigma^*$ & $\Xi^*$ & $\Omega^-$ \\
     \hline
    $\xi_{D,\pi}$ & $t_0+3t_D$ & $t_0+t_D$ & $t_0-t_D$ & $t_0-3t_D$ \\
    $\xi_{D,K}$ & $2t_0$ & $2t_0+2t_D$ & $2t_0+4t_D$ & $2t_0+6t_D$ \\
    \hline\hline
  \end{tabular}

\end{table}

The octet baryon masses up to N$^3$LO  and with the virtual decuplet contributions can be written as
\begin{equation}\label{masseq}
  m_B = m_0 + m_B^{(2)} + m_B^{(3)} + m_B^{(4)} + m_B^{(D)}.
\end{equation}
Here, $m_0$ is the chiral limit octet-baryon mass, and $m_B^{(2)}$, $m_B^{(3)}$, $m_B^{(4)}$ correspond to the $\mathcal{O}(p^2)$, $\mathcal{O}(p^3)$, $\mathcal{O}(p^4)$ contributions from the octet-only EOMS BChPT, respectively. Their explicit expressions can be found in Ref.~\cite{Ren:2012aj}, where the PCB terms are already removed.
The last term $m_B^{(D)}$ denotes the contributions of the virtual decuplet resonances up to N$^3$LO.
After calculating the Feynman diagrams shown in Fig.~\ref{Fig1:FeynmanD} and subtracting the PCB terms with the EOMS scheme, the virtual decuplet contributions to the octet baryon masses can be expressed as
\begin{equation}\label{masseqD}
  m_B^{(D)}=\frac{1}{(4\pi F_{\phi})^2}\sum\limits_{\phi=\pi,~K,~\eta}\xi_{B,\phi}^{(a)}H_B^{(a)}(M_{\phi}) + \frac{1}{(4\pi F_{\phi})^2}\sum_{\substack{\phi=\pi,~K,~\eta\\  D=\Delta,~\Sigma^*,~\Xi^*,~\Omega^-}}\xi_{BD,\phi}^{(b/c)}\cdot H_{B,D}^{(b/c)}(M_{\phi}).
\end{equation}
The first term of Eq.~(\ref{masseqD}) is the NNLO contributions of Feynman diagram Fig.~\ref{Fig1:FeynmanD}(a). The loop function $H_B^{(a)}(M_{\phi})$ can be found in Ref.~\cite{MartinCamalich:2010fp} and the corresponding coefficients $\xi_{B,\phi}^{(a)}$ are listed in Table~\ref{Coef:nlo}. The next term is the virtual decuplet contribution at $\mathcal{O}(p^4)$ from the one-loop diagram of Fig.~\ref{Fig1:FeynmanD}(b) and the wave function renormalization diagrams of Fig.~\ref{Fig1:FeynmanD}(c). The Clebsch-Gordan coefficients $\xi_{BD,\phi}^{(b/c)}$ are tabulated in Table~\ref{Coef:nnlo}, and the loop function  $H_B^{(b/c)}(M_{\phi})$ has the following form:
{
\begin{eqnarray}
  H_B^{(b/c)}(M_{\phi}) &&= \frac{1}{24m_0^2m_D^2}m_B^{(2)}M_{\phi}^2 \left[2(11m_0^2+8m_0m_D+9m_D^2)\right.\nonumber\\
  &&\qquad \left. \times (m_0^2-m_D^2) + (5m_0^2+8m_0m_D+18m_D^2)M_{\phi}^2-6M_{\phi}^4\right]\nonumber\\
  &&\quad + \frac{1}{36m_0m_D^3}m_D^{(2)}M_{\phi}^2 \left[-34m_0^4-24m_0^3m_D+30m_D^4\right.\nonumber\\
  &&\qquad\left. -6m_D^2M_{\phi}^2-6M_{\phi}^4 +6m_0m_D(4m_D^2+M_{\phi}^2)+3m_0^2(4m_D^2+7M_{\phi}^2)\right]\nonumber\\
  && \quad - \frac{1}{12m_0^4m_D^2}m_B^{(2)}M_{\phi}^2\ln\left(\frac{M_{\phi}}{m_D}\right) \left[12m_D^5(m_0+m_D)\right. \nonumber\\
  &&\qquad \left.+6(m_0^4-m_0^2m_D^2-2m_0m_D^3-3m_D^4)M_{\phi}^2+ 4(m_0^2+m_0m_D+3m_D^2)M_{\phi}^4 - 3M_{\phi}^6\right]\nonumber\\
  && \quad + \frac{1}{6m_0^3m_D^3}m_D^{(2)}M_{\phi}^2\ln\left(\frac{M_{\phi}}{m_D}\right) \left[m_D^5(9m_0+8m_D)\right.\nonumber\\
  &&\qquad \left.+3(m_0^2-m_D^2)(2m_0^2+m_0m_D+2m_D^2)M_{\phi}^2-m_0(4m_0+m_D)M_{\phi}^4 + M_{\phi}^6\right]\nonumber\\
  &&\quad - \frac{1}{12m_0^4m_D^3} M_{\phi}^2 (m_0-m_D)^2(m_0+m_D)^4\ln\left(\frac{m_DM_{\phi}}{m_D^2-m_0^2}\right)\nonumber\\
  &&\qquad \times\left[m_D\left(-5m_0^2+2m_0m_D-3m_D^2\right)m_B^{(2)} + 2m_0\left(m_0^2-m_0m_D+3m_D^2\right)m_D^{(2)}\right]\nonumber\\
  &&\quad - \frac{1}{6m_D^3}m_0M_{\phi}^4\ln\left(\frac{m_DM_{\phi}}{\mu^2}\right)\left[6m_D(m_0+m_D)m_{B}^{(2)} - m_0(4m_0+3m_D)m_D^{(2)}\right]\nonumber\\
  &&\quad + \frac{1}{12m_0^4m_D^3\sqrt{\mathcal{W}}} (m_0^2-2m_0m_D+m_D^2-M_{\phi}^2)(m_0^2+2m_0m_D+m_D^2-M_{\phi}^2)^2\nonumber\\
  &&\qquad \times \left[3m_D(m_D^2-M_{\phi}^2)^2m_{B}^{(2)}-2m_0(m_D^2-M_{\phi}^2) \left((3m_D^2+M_{\phi}^2)m_D^{(2)}+m_D^2m_B^{(2)}\right)\right.\nonumber\\
  &&\qquad\quad + 2m_0^2m_D^2 (m_D^2+M_{\phi}^2)(m_B^{(2)}+m_D^{(2)})+ 2m_0^3\left(2(m_D^2-M_{\phi}^2)m_D^{(2)}+m_D^2m_B^{(2)}\right)\nonumber\\
  &&\qquad\quad \left.-m_0^4m_D\left(5m_B^{(2)}+2m_D^{(2)}\right)+2m_0^5m_D^{(2)}\right]\nonumber\\
  &&\qquad \times \left[\arctan{\left(\frac{m_0^2+m_D^2-M_{\phi}^2}{\mathcal{W}}\right)} +\arctan{\left(\frac{m_0^2-m_D^2+M_{\phi}^2}{\mathcal{W}}\right)}\right],
\end{eqnarray}
}
where $\mathcal{W}=-m_0^4-(m_D^2-M_{\phi}^2)^2+2m_0^2(m_D^2+M_{\phi}^2)$,
the NLO octet baryon mass $m_B^{(2)}$  is given in Eq.~(\ref{masseq}) and the NLO decuplet baryon mass is
 \begin{equation}\label{mass:D2}
 m_D^{(2)} = - \sum\limits_{\phi=\pi,K}\xi_{D,\phi}^{(2)}M_{\phi}^2.
 \end{equation}
 The corresponding coefficients $\xi_{D,\phi}^{(2)}$ are listed in Table~\ref{Coef:dlo}. 
It should be noted
that in obtaining the results of Eq.~(\ref{masseqD}), the decuplet-octet mass difference, $\delta=m_D-m_0$, is considered up to all orders~\cite{MartinCamalich:2010fp}.

\subsection{Finite-volume corrections from the virtual deucplet baryons}

To study the LQCD data from the PACS-CS~\cite{Aoki:2008sm}, LHPC~\cite{WalkerLoud:2008bp}, HSC~\cite{Lin:2008pr}, QCDSF-UKQCD~\cite{Bietenholz:2011qq} and NPLQCD~\cite{Beane:2011pc} collaborations, FVCs should be taken into account. That's because at present all LQCD simulations are performed in a finite hypercube with lattice sizes $L\sim 3-5$ fm and the finite-volume effects on the simulation results cannot be neglected. Recently, the FVCs to the octet baryon masses have been  studied in the HBChPT~\cite{Beane:2004tw} and the EOMS BChPT~\cite{Geng:2011wq} up to NNLO. It was shown that the FVCs can be helpful to constrain some relevant LECs. In order to self-consistently take into account the FVCs  to the octet baryon masses, we need to calculate the FVCs up to N$^3$LO in the EOMS BChPT. Following the same procedure as detailed in Ref.~\cite{Geng:2011wq}, one can easily calculate the $\mathcal{O}(p^4)$ BChPT results in a finite hypercube by replacing the $H$ of Eq.~(\ref{masseq}) with $\tilde{H}=H+\delta G$. For the purpose of this work, we only present the FVCs from the virtual decuplet baryons. The octet-only formulas of FVCs can be found in Ref.~\cite{Geng:2011wq,Ren:2012aj}.

The FVCs to the loop results of  Fig.~\ref{Fig1:FeynmanD}(a,b,c) are
\begin{equation}
  \delta G_B^{(a)} = \frac{3}{4} \int_0^1 dx \left[\frac{m_0^2\left(m_0(1-x)+m_D\right)}{6m_D^2}\delta_{1/2}\left(\mathcal{M}_D^2\right) -\frac{m_0^2\left(m_0(1-x)+m_D\right)\mathcal{M}_D^2}{6m_D^2}\delta_{3/2}\left(\mathcal{M}_D^2\right)\right]
\end{equation}
and
\begin{eqnarray}
  \delta G_{B,D}^{(b/c)} &&=\frac{m_0}{6m_D^3}\int_0^1 dx \left\{
  \left[2m_0^2(x-1)m_D^{(2)}-m_0m_D\left(m_D^{(2)}+3m_B^{(2)}(x-1)\right)+2m_D^2m_B^{(2)}\right]\times
  \delta_{1/2}\left(\mathcal{M}_D^2\right)\right. \nonumber\\
  &&\quad +\left[3m_0^3m_Dx(x-1)^2m_B^{(2)}-2m_0^2\mathcal{M}_D^2(x-1)m_D^{(2)} + 3m_0^2m_D^2x(x-1)\left(m_D^{(2)}-m_B^{(2)}\right)\right.\nonumber\\
  &&\qquad \left.+m_0m_D\mathcal{M}_D^2\left(3m_B^{(2)}(x-1)+m_D^{(2)}\right)-3m_0m_D^3m_D^{(2)}x - 2m_D^2\mathcal{M}_D^2m_B^{(2)}\right]\times\delta_{3/2}\left(\mathcal{M}_D^2\right)\nonumber\\
  &&\quad \left. +3m_0m_D\mathcal{M}_D^2x\left[-m_0^2(x-1)^2m_B^{(2)} +m_0m_D(x-1)\left(m_B^{(2)}-m_D^{(2)}\right)+m_D^2m_D^{(2)}\right]\nonumber \right.\\
  &&\left.\hspace{3cm}\times \delta_{5/2}\left(\mathcal{M}_D^2\right)\right\},
\end{eqnarray}
respectively. Here $\mathcal{M}_D^2=x^2m_0^2- x(m_0^2-m_D^2) + (1-x)M_{\phi}^2-i\epsilon$, and the ``master" formulas are defined as
\begin{equation}
\delta_r (\mathcal{M}^2)=
\frac{2^{-1/2-r}(\sqrt{\mathcal{M}^2})^{3-2r}}{\pi^{3/2}\Gamma(r)}
\sum_{\vec{n}\ne0}(L\sqrt{\mathcal{M}^2}|\vec{n}|)^{-3/2+r}K_{3/2-r}(L\sqrt{\mathcal{M}^2}|\vec{n}|),
\end{equation}
where $K_n(z)$ is the modified Bessel function of the second kind, and $\sum\limits_{\vec{n}\ne0}\equiv\sum\limits^{\infty}_{n_x=-\infty}\sum\limits^{\infty}_{n_y=-\infty}\sum\limits^{\infty}_{n_z=-\infty}(1-\delta(|\vec{n}|,0))$ with $\vec{n}=(n_x,n_y,n_z)$.

\section{Results and Discussions}\label{SecIII}
In this section, the effects of the virtual decuplet baryons on the g.s. octet baryon masses are systematically studied by fitting the $n_f=2+1$ LQCD data of the PACS-CS~\cite{Aoki:2008sm}, LHPC~\cite{WalkerLoud:2008bp}, HSC~\cite{Lin:2008pr}, QCDSF-UKQCD~\cite{Bietenholz:2011qq} and NPLQCD~\cite{Beane:2011pc} collaborations.

\subsection{Light-quark mass dependence of the octet baryon masses}

Up to N$^3$LO, there are $19$ unknown LECs ($m_0$, $b_0$, $b_D$, $b_F$, $b_{1-8}$, and $d_{1-5,~7,~8}$) in the octet-only EOMS BChPT.
To take into account the contributions of the decuplet baryons,
one has to introduce four more LECs, $m_D$, $t_0$, $t_D$ and $\mathcal{C}$ (see Section IIA).  The $\phi B D$ coupling constant $\mathcal{C}$ can be fixed to
the SU(3)-average value among the different decuplet-to-octet pionic decay channels, i.e., $\mathcal{C}=0.85$~\cite{Alarcon:2012nr}~\footnote{In Refs.~\cite{Geng:2009hh,MartinCamalich:2010fp} the value of $\mathcal{C}$ is fixed from the $\Delta(1232)\rightarrow\pi N$ decay rate, which yields $\mathcal{C}=1.0$. But in our previous study of the NPLQCD data, this coupling turned out to be somewhat smaller~\cite{Geng:2011wq}.}. A moderate variation of $\mathcal{C}$ has no significant effects on our final results. The LECs $t_0$, $t_D$, and $m_D$
can be fixed by fitting the NLO decuplet mass formula $M_D = m_D-m_D^{(2)}$ to the physical decuplet baryon masses. Because $t_0$ and $m_D$ cannot be disentangled at the physical point, one only obtains a combination of $m_D$ and $t_0$ with $m_D^{\rm eff}=m_D-t_0(2M_K^2+M_{\pi}^2)=1.215$ GeV and $t_D=-0.326$ GeV$^{-1}$. In the following, the octet-decuplet mass splitting $\delta=m_D-m_0$ is fixed to be $0.231$ GeV--the average mass gap between the octet and decuplet baryons.
Therefore, one can fix the four LECs in the following way: $m_D=m_0+0.231$ GeV, $t_0=(m_0-0.984)/0.507$ GeV$^{-1}$, $t_D=-0.326$ GeV$^{-1}$ and $\mathcal{C}=0.85$.
As a result, the same $19$ LECs as those in the octet-only BChPT need to be determined. The other coupling constants are fixed as in Ref.~\cite{Ren:2012aj}: the meson decay constant $F_{\phi}=0.0871$ GeV, the baryon axial coupling constants $D=0.8$, $F=0.46$ and the renormalization scale $\mu=1$ GeV.

Following the same procedure as in Ref.~\cite{Ren:2012aj}, we use the formulas Eq.~(\ref{masseq}) to fit the octet baryon masses of the LQCD simulations and of their experimental values~\cite{Beringer:1900zz}. The LQCD data to be studied are taken from the PACS-CS~\cite{Aoki:2008sm}, LHPC~\cite{WalkerLoud:2008bp}, HSC~\cite{Lin:2008pr}, QCDSF-UKQCD~\cite{Bietenholz:2011qq} and NPLQCD~\cite{Beane:2011pc}  data satisfying $M_{\pi}<500$ MeV and $M_{\pi}L>4$, named as Set-I in Ref.~\cite{Ren:2012aj}. The restrictions on the lattice data
are taken to ensure that the N$^3$LO BChPT is valid for these pion (light-quark) masses and lattice volumes. Later we will slightly relax the restrictions to test whether the inclusion of the virtual decuplet baryons can extend the applicability region of the BChPT.
The so-obtained values of the LECs from the best fit and the corresponding $\chi^2/\mathrm{d.o.f.}$ are tabulated in Table~\ref{Tb:fitcoef}.
For comparison, we also list the octet-only best fit results of lattice data Set-I from Ref.~\cite{Ren:2012aj}.
From the $\chi^2/\mathrm{d.o.f.}$, one can conclude that the inclusion of the virtual decuplet baryons
does not change the description of the LQCD data. On the other hand,  the values of the LECs have changed a lot, as can be clearly seen from Table~\ref{Tb:fitcoef}~\footnote{The same phenomenon has been observed in the studies of the octet baryon magnetic moments~\cite{Geng:2009hh} and the octet baryon masses up to NNLO~\cite{Alarcon:2012nr}.}. This confirms the assumption that using only the octet baryon mass data, one can not disentangle the virtual decuplet contributions from those of the virtual octet baryons and the tree-level diagrams~\cite{Ren:2012aj}. In other words, for the static properties of the octet baryons, most contributions of the virtual decuplet are hidden in the relevant LECs, as one naively expects. Below, we will see that 
their inclusion, however, does improve the description of the volume-dependence of the LQCD data, as also noted in Ref.~\cite{Geng:2011wq}.

{\small
\begin{table}[t]
\centering
\caption{Values of the LECs from the best fit to the LQCD data and the experimental data at
 $\mathcal{O}(p^4)$ (see text for details).}
\label{Tb:fitcoef}
\begin{tabular}{l|rr|r|r}
\hline\hline
       & \multicolumn{2}{c|}{Fit-I}   & Fit-12      & Fit-13 \\
       \cline{2-3}
       &   w/o decuplet & w/ decuplet & w/ decuplet  &  w/ decuplet \\
\hline
  $m_0$~[MeV]         & $879(22)$      & $908(24)$       & $910(17)$      &   $910(23)$    \\
  $b_0$~[GeV$^{-1}$]  & $-0.609(19)$   & $-0.744(16)$    &$-0.757(13)$    &  $-0.772(16)$  \\
  $b_D$~[GeV$^{-1}$]  & $0.225(34)$    & $0.355(20)$     &$0.352(21)$     &  $0.355(20)$   \\
  $b_F$~[GeV$^{-1}$]  & $-0.404(27)$   & $-0.552(28)$    &$-0.548(23)$    &  $-0.555(26)$  \\
  \hline
  $b_1$~[GeV$^{-1}$]  & $0.550(44)$    & $1.08(6)$       & $1.39(7)$      &  $1.52(8)$    \\
    $b_2$~[GeV$^{-1}$]  & $-0.706(99)$ & $0.431(93)$     & $0.650(72)$   &  $0.555(138)$    \\
  $b_3$~[GeV$^{-1}$]  & $-0.674(115)$    & $-1.83(15)$     & $-1.07(13)$    &  $-1.95(17)$    \\
  $b_4$~[GeV$^{-1}$]  & $-0.843(81)$   & $-1.57(4)$      & $-1.67(2)$    &  $-1.59(6)$    \\
  $b_5$~[GeV$^{-2}$]  & $-0.555(144)$  & $-0.355(74)$    & $-1.03(9)$  &  $-1.32(2)$   \\
  $b_6$~[GeV$^{-2}$]  & $0.160(95)$    & $-0.423(117)$   & $0.115(95)$   &  $-0.297(35)$   \\
  $b_7$~[GeV$^{-2}$]  & $1.98(18)$     & $2.79(15)$      & $2.32(11)$     &  $2.63(5)$     \\
  $b_8$~[GeV$^{-2}$]  & $0.473(65)$    & $-1.73(6)$      & $-1.62(3)$    &  $-1.96(2)$    \\
  \hline
  $d_1$~[GeV$^{-3}$]  & $0.0340(143)$  & $0.0157(130)$   & $0.00416(1296)$   &  $-0.00418(1330)$  \\
  $d_2$~[GeV$^{-3}$]  & $0.296(53)$    & $0.445(57)$     & $0.441(49)$    &  $0.464(55)$     \\
  $d_3$~[GeV$^{-3}$]  & $0.0431(304)$  & $0.328(18)$     & $0.280(22)$    &  $0.264(15)$     \\
  $d_4$~[GeV$^{-3}$]  & $0.234(67)$    & $-0.117(59)$    & $-0.0345(676)$  &  $0.00590(5393)$   \\
  $d_5$~[GeV$^{-3}$]  & $-0.328(60)$   & $-0.853(77)$    & $-0.831(59)$   &  $-0.883(69)$     \\
  $d_7$~[GeV$^{-3}$]  & $-0.0358(269)$ & $-0.425(39)$    & $-0.464(12)$   &  $-0.497(33)$     \\
  $d_8$~[GeV$^{-3}$]  & $-0.107(32)$   & $-0.557(56)$    & $-0.602(18)$   &  $-0.651(50)$      \\
  \hline
$\chi^2/\mathrm{d.o.f.}$       & $1.0$         & $1.0$          & $1.0$     & $1.2$  \\
\hline\hline
\end{tabular}

\end{table}
}

\begin{figure}[t]
  \centering
  \includegraphics[width=14cm]{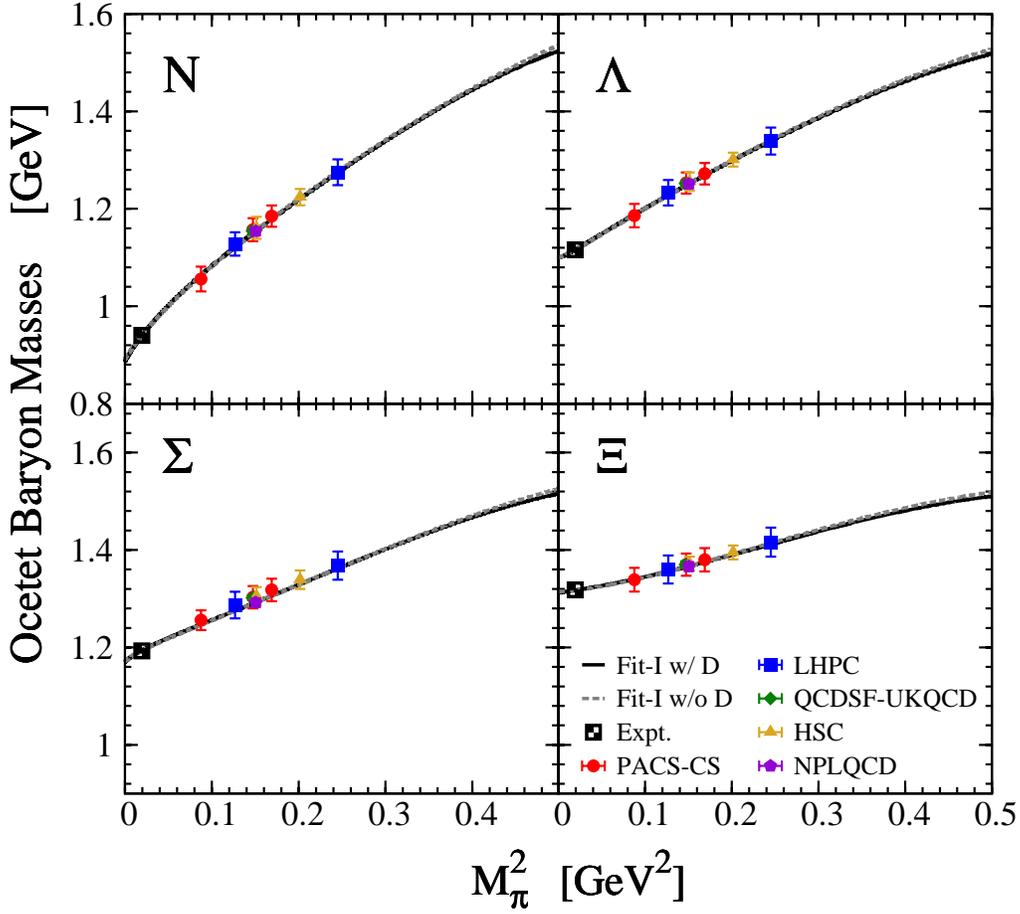}\\
  \caption{(Color online). Pion mass dependence of the LQCD data in comparison with the best fits of the EOMS BChPT
  up to N$^3$LO with (solid lines) and without (dashed lines) the virtual decuplet contributions. The lattice data have been extrapolated to
   the physical strange-quark mass and infinite space-time.}
  \label{Re:chiex}
\end{figure}

In Fig.~\ref{Re:chiex}, setting the strange-quark mass to its physical value, we show the pion mass dependence of the octet baryon masses in the N$^3$LO EOMS BChPT with and without the virtual decuplet baryon contributions. It is clear that the two N$^3$LO fits give the same description of lattice data Set-I, as can be inferred from the same $\chi^2/\mathrm{d.o.f.}$ shown in Table \ref{Tb:fitcoef}.

\subsection{Finite-volume corrections to the octet baryon masses}

In Ref.~\cite{Geng:2011wq}, we have studied the FVCs to the g.s. octet baryon masses using the EOMS BChPT up to NNLO, and found that the finite-volume effects are very important and cannot be neglected. Therefore, in this work the FVCs are self-consistently included in Eq.~(\ref{masseq}) to analyze the lattice data. The NPLQCD~\cite{Beane:2011pc} simulation is performed with the same pion mass of $M_{\pi}\simeq 390$ MeV and at four different lattice sizes $L\sim 2.0,~2.5,~3.0$ and $3.9$ fm. Therefore, it provides a good opportunity to study the FVCs to the octet-baryon masses.

\begin{figure}[b]
  \centering
  \includegraphics[width=15cm]{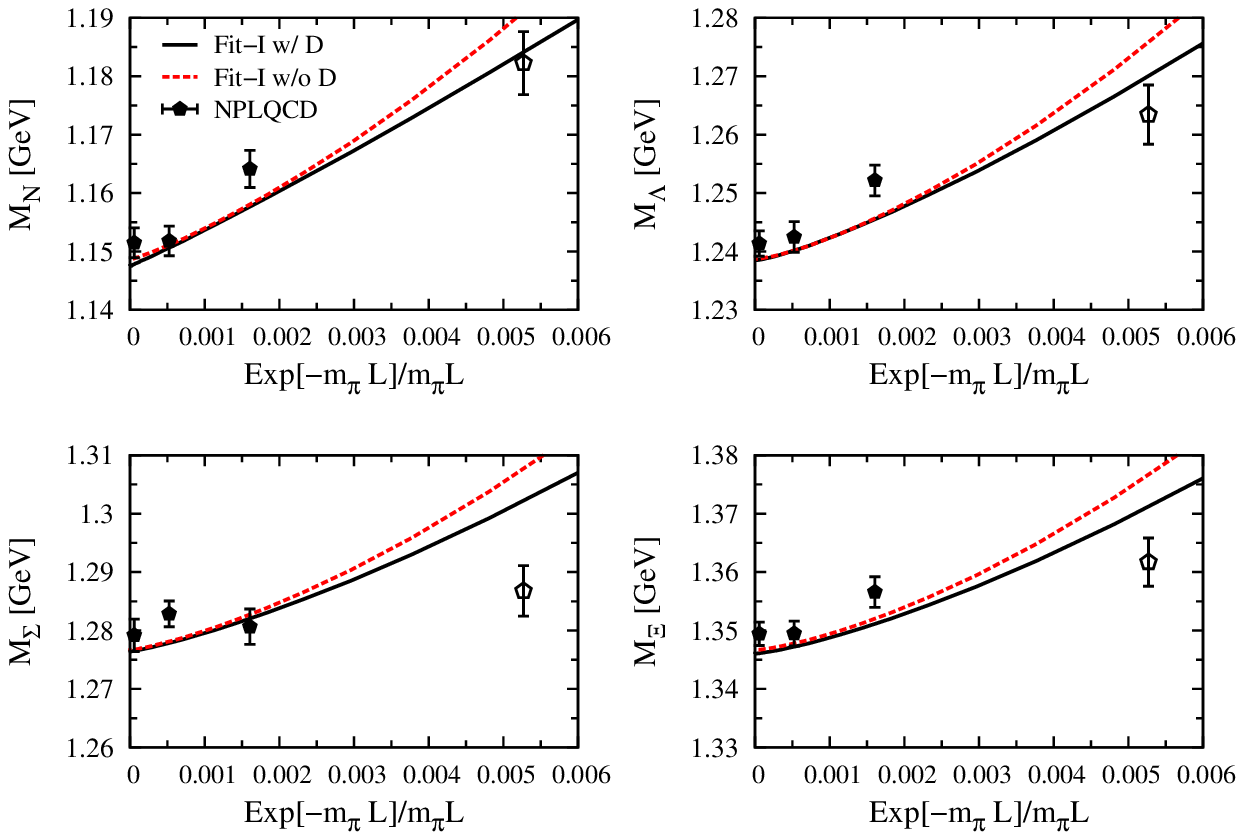}\\
  \caption{(Color online). Lattice volume dependence of the NPLQCD data in comparison with the EOMS BChPT up to N$^3$LO with (solid lines) and without (dashed lines) the virtual decuplet contributions. The three black points with $M_{\phi}L>4$ are included in data Set-I, while the hollow points with $M_{\phi}L=3.86$ are not.}
  \label{Re:fvc}
\end{figure}

 In Fig.~\ref{Re:fvc}, we contrast the NPLQCD data with the N$^3$LO EOMS BChPT using the Fit-I LECs from Table \ref{Tb:fitcoef}. As stated in Ref.~\cite{Ren:2012aj}, three sets of the NPLQCD data with $M_{\phi}L>4$ are included in lattice data Set-I and denoted by solid points in Fig.~\ref{Re:fvc}. Another set with $M_{\pi}L=3.86$ (hollow points) is not included. Both the octet-only and the octet plus decuplet (O+D) BChPT can give a reasonable description of the FVCs. At the $e^{-m_{\pi}L}/(m_{\pi}L)\leq0.2$ region, these two fits give essentially the same results. With the increase of $e^{-m_{\pi}L}/(m_{\pi}L)$ ( the decrease of lattice size $L$), the O+D BChPT results are in better agreement the NPLQCD data, especially for the nucleon mass. It seems that the virtual decuplet baryons can help to improve the description of the FVCs, although the BChPT results are still a bit larger than the LQCD data at small $M_{\phi}L$.

\begin{figure}
  \centering
  \includegraphics[width=15cm]{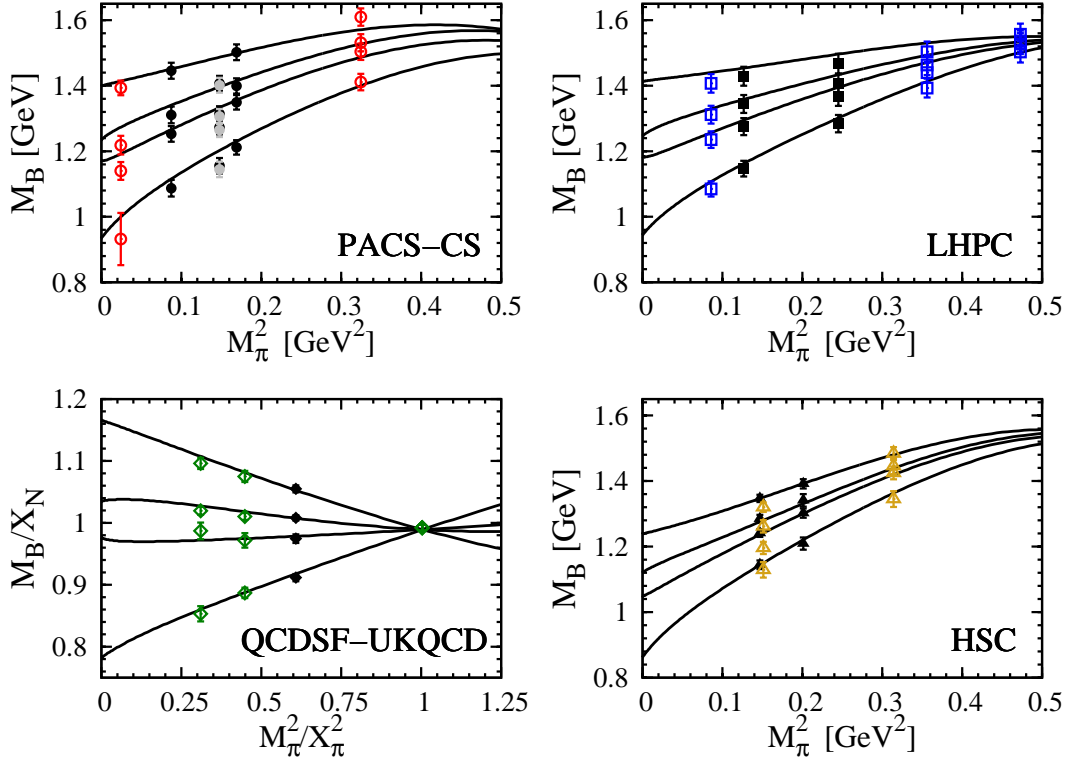}\\
  \caption{(Color online). The PACS-CS, LHPC, QCDSF-UKQCD and HSC lattice data in comparison with the O+D BChPT best fit as functions of the pion mass. The lines in each panel (from bottom to top) correspond to $N$, $\Lambda$, $\Sigma$ and $\Xi$, respectively. The kaon mass is fixed using $M_K^2=a+bM_{\pi}^2$ for the corresponding lattice ensemble with $a$ and $b$ determined in Ref.~\cite{Ren:2012aj}. The lattice data have been extrapolated to infinite space-time using the corresponding BChPT fit. $X_{\pi}=\sqrt{(M_{\pi}^2+2M_K^2)/3}$, $X_N=(m_N+m_{\Sigma}+m_{\Xi})/3$, where the meson and baryon masses are the physical ones.}
  \label{Re:description}
\end{figure}

It is interesting to check whether the O+D best fit can describe the lattice data with larger pion masses and/or smaller lattice volumes.
In Fig.~\ref{Re:description},  the PACS-CS, LHPC, HSC and QCDSF-UKQCD lattice data with $M_{\pi}<700$ MeV are
compared with the best N$^3$LO O+D EOMS BChPT with the Fit-I LECs of Table~\ref{Tb:fitcoef}. The lattice points  included in the fit are denoted by solid points and those excluded in the fit by hollow points. It is clear that
the N$^3$LO BChPT can describe reasonably well the LQCD data, even those excluded in
the fit. The average deviation of the BChPT results from the LQCD data, defined as~\footnote{The uncertainty of the lattice data, $\Delta^i_\mathrm{LQCD}$, can be found in Ref.~\cite{Ren:2012aj}.}
$$
 \tilde{\chi}^2=\frac{1}{N_\mathrm{LQCD}}\sum\limits_{i=1}^{N_\mathrm{LQCD}} \left(\frac{M^i_\mathrm{LQCD}-M^i_\mathrm{BChPT}}{\Delta^i_\mathrm{LQCD}}\right)^2,
$$
is $3.1$, $2.5$, $1.2$ and $1.2$ for the PACS-CS, LHPC, HSC, and QCDSF-UKQCD data, respectively. Here, it should be noted that in Fig.~\ref{Re:description}, only the QCDSF-UKQCD data with $N_s=32$ are shown and those simulated in a smaller volume with $N_s=24$ are not explicitly displayed. Including them in the $\tilde{\chi}^2$, one would have obtained a $\tilde{\chi}^2=22.3$.

It is clear from the above comparisons that using the LECs determined from the best fit to lattice data Set-I, the BChPT cannot well describe the LQCD data obtained in smaller volumes, particularly those of the QCSDSF-UKQCD data with $N_s=24$. On the other hand, the virtual decuplet contributions seem to be helpful in this regard. Furthermore, it should be noted that we have chosen lattice data set-I by requiring $M_\pi<500 $ MeV and $M_\pi L>4$. These criteria yielded a $\chi^2/\mathrm{d.o.f.}=1$, but nevertheless, are a bit arbitrary. In the following subsection, we would like to slightly relax the above criteria and study whether the LQCD data with smaller $M_\pi L$ can be described at a reasonable sacrifice of the $\chi^2/\mathrm{d.o.f.}$.

\subsection{Description of the NPLQCD and QCDSF-UKQCD small-volume data}

In order to better describe the LQCD data obtained in smaller volumes, we have to take into account in the fit
 a few more LQCD data with small $M_{\phi}L$. Furthermore, to guarantee that the N$^3$LO BChPT is still
 valid, we have to ensure that the corresponding $\chi^2/\mathrm{d.o.f.}$ be around $1$.
 Therefore, we slightly relax the criterium of $M_{\phi}L>4$ to $M_{\phi}L>3.8$ and keep that of $M_{\pi}<500$ MeV. Fourteen sets of LQCD data from  the PACS-CS, LHPC, HSC, QCDSF-UKQCD and NPLQCD collaborations satisfy the new criteria~\cite{Ren:2012aj}.

\begin{figure}[t]
  \centering
  \includegraphics[width=15cm]{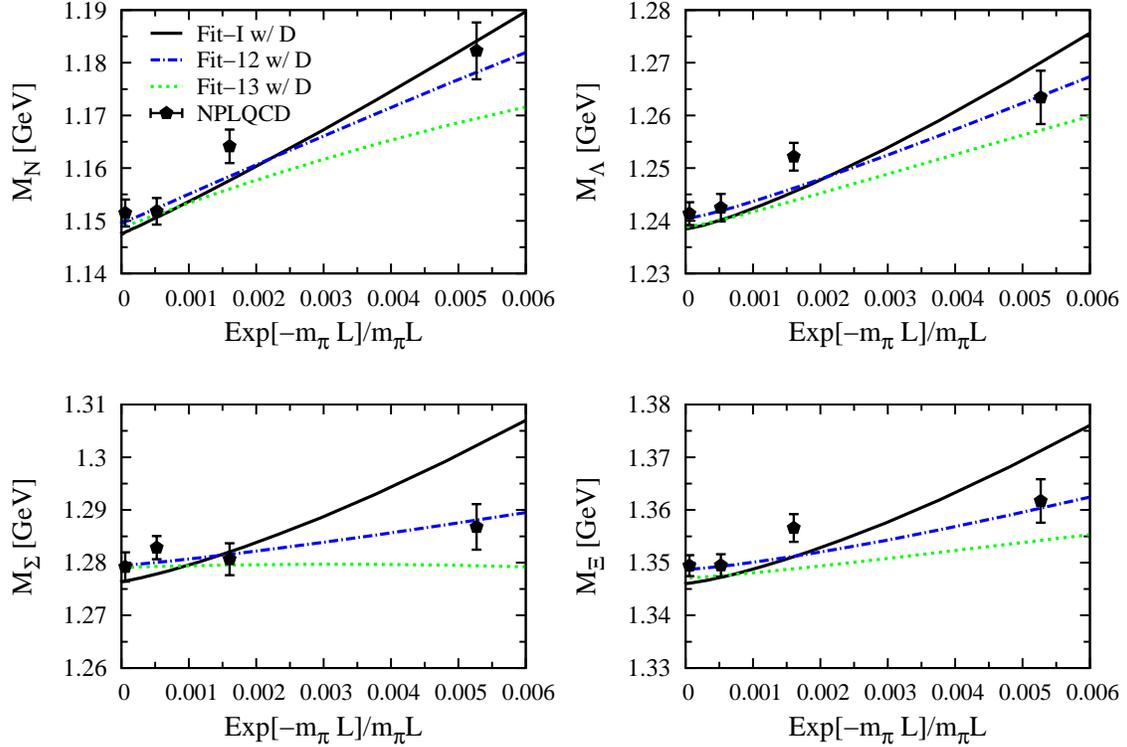}\\
  \caption{(Color online). Lattice volume dependence of the NPLQCD data in comparison with the best fits of the N$^3$LO O+D BChPT. The solid lines correspond to the results of Fit-I, the blue-dot-dashed lines to the results of Fit-12 and the green-hashed lines denote the results of Fit-13.}
  \label{Re:fvc1213}
\end{figure}

\begin{figure}[t]
  \centering
  \includegraphics[width=15cm]{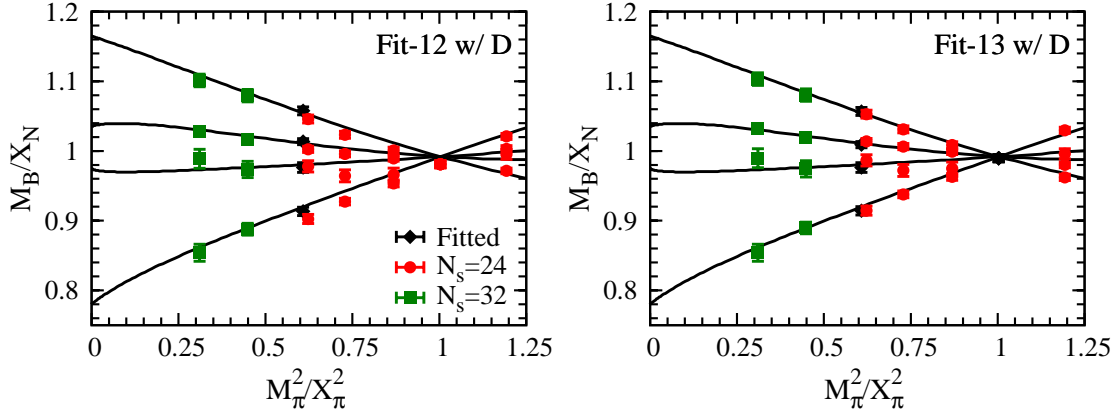}\\
  \caption{The QCDSF-UKQCD lattice data in comparison with the N$^3$LO O+D BChPT results of Fit-12 (left panel) and Fit-13 (right panel). The black points are included in the data Set-12/13, the red (with $N_s=24$) and blue (with $N_s=32$) points are not. The FVCs have been subtracted from the lattice data using the corresponding BChPT fit.}
  \label{Re:desUKQCD}
\end{figure}

Because the NPLQCD results ~\cite{Beane:2011pc} are obtained with the same light-quark masses but at different lattice sizes, they
are more suitable for studies of FVCs and can better constrain the relevant LECs. First, we add the NPLQCD data point with $M_{\phi}L=3.86$ to lattice data Set-I to form a new data set--Set-12, and fit it using the N$^3$LO O+D BChPT. In Table~\ref{Tb:fitcoef}, the LECs from
the best fit (named Fit-12) and the corresponding $\chi^2/\mathrm{d.o.f.}=1.0$ are listed~\footnote{If the octet-only BChPT were
used to fit data Set-12, the best $\chi^2/\mathrm{d.o.f.}$ would be $1.1$.}. Compared with the O+D Fit-I results, most of the LECs do not change much, particularly those of $b_{0}$, $b_D$, $b_F$ and $m_0$. In Fig.~\ref{Re:fvc1213}, the finite-volume dependences of the NPLQCD data are compared
 with the N$^3$LO O+D BChPT results calculated using the Fit-12 LECs. The BChPT results are in very good agreement with the lattice data and the corresponding $\tilde{\chi}^2$ is $0.7$. It seems that the Fit-12 LECs can give a better description of the LQCD data
 with small $M_\phi L$.  This is further confirmed by contrasting the Fit-12 results to the QCDSF-UKQCD lattice data. As shown in Fig.~\ref{Re:desUKQCD},  both the data with $N_s=32$ (green points) and $N_s=24$ (red points) are better described than the BChPT results using the Fit-I LECs. The corresponding $\tilde{\chi}^2$ are 0.6 and 7.3, respectively.

 We note that the NPLQCD simulation is performed at fixed $M_{\pi}\simeq390$ MeV and close to the physical strange-quark mass,
 and the values of $M_{K}L$,  $M_{\eta}L$ are all larger than $5$ (for the smallest lattice size $L=2.0$ fm, $M_K L=5.4$ and $M_{\eta}L=5.8$). On the other hand, the situation is different for the QCDSF-UKQCD collaboration~\cite{Bietenholz:2011qq}. They start with a common sea quark mass near $(2m_{ud}^{\rm phys.}+ m_s^{\rm phys.})/3$ and split the masses symmetrically~\cite{Durr:2013qk}. Thus, for the data with $N_s=24$, not only $M_{\pi}L$ is smaller than $4$, but $M_{K}L$ and $M_{\eta}L$ are all around $4$.

In order to take into account the finite-volume effects induced by the smaller $M_{K}L$ and $M_{\eta}L$,
we add to data Set-12 the LQCD data set with $3.8<M_{\phi}L<4$ from the QCDSF-UKQCD collaboration. The new data set--Set-13--has $13$ lattice data points. Using the N$^3$LO O+D BChPT to fit Set-13, we obtain the results tabulated in Table~\ref{Tb:fitcoef} (named as Fit-13). Compared to Fit-12, the values of the LECs do not change dramatically, and the corresponding $\chi^2/\mathrm{d.o.f.}$ becomes $1.2$~\footnote{If the octet-only BChPT were used, the obtained $\chi^2/\mathrm{d.o.f.}$ would be $1.7$.}. In Fig.~\ref{Re:fvc1213} and Fig.~\ref{Re:desUKQCD}, we show the N$^3$LO O+D BChPT results calculated using the Fit-13 LECs. It is clear that the description of the NPLQCD data becomes a bit worse ($\tilde{\chi}^2=1.1$),
while the QCDSF-UKQCD data can be better described, with $\tilde{\chi}^2=0.6$ for $N_s=32$ and $\tilde{\chi}^2=4.2$ for $N_s=24$ lattice data, respectively. Nevertheless, it seems that up to N$^3$LO, the O+D EOMS BChPT still cannot achieve a perfect description of the QCDSF-UKQCD data with $N_s=24$.

It should be pointed out that because Fit-12 can describe
better the volume dependence of the NPLQCD and QCDSF-UKQCD data while still yields a $\chi^2/\mathrm{d.o.f.}=1.0$, it is preferable than Fit-13 and Fit-I.

\subsection{Pion- and strangeness-baryon sigma terms}

\begin{table}[b]
\centering
\caption{Sigma terms of the octet baryons predicted by the N$^3$LO BChPT with
 LECs determined from the global fits to different lattice data sets. The first error is statistical and the second is systematic, estimated by taking half the difference between the N$^3$LO result and the NNLO result.}
\label{sigmaresults}
\begin{tabular}{c|rr|r|r||rr|r|r}
\hline\hline
       & \multicolumn{4}{c||}{$\sigma_{\pi B}$}        & \multicolumn{4}{c}{$\sigma_{sB}$}\\
\cline{2-5} \cline{6-9}
       & \multicolumn{2}{c|}{Fit-I}  & Fit-12 & Fit-13 & \multicolumn{2}{c|}{Fit-I}  & Fit-12 & Fit-13\\
\cline{2-3} \cline{6-7}
       &  w/o decuplet & w/ decuplet & w/decuplet & w/decuplet & w/o decuplet & w/ decuplet & w/decuplet & w/decuplet\\
\hline
$N$       &  $43(1)(6)$ & $46(2)(12)$ & $47(1)(12)$ & $47(2)(12)$ & $126(24)(54)$ & $157(25)(68)$ & $149(22)(63)$ & $157(22)(61)$   \\
$\Lambda$ &  $19(1)(7)$ & $20(2)(13)$ & $21(2)(13)$ & $22(2)(12)$ & $269(23)(66)$ & $256(22)(60)$ & $250(25)(54)$ & $257(19)(51)$  \\
$\Sigma$  &  $18(2)(6)$ & $19(2)(6)$ & $20(2)(7)$ & $21(2)(6)$ & $296(21)(50)$ & $270(22)(47)$  & $266(23)(46)$ & $272(20)(50)$  \\
$\Xi$     &  $4(2)(3) $ & $6(2)(5) $ & $7(2)(4) $ & $7(2)(5) $ & $397(22)(56)$ & $369(23)(50)$  & $366(23)(48)$ & $372(21)(49)$  \\
\hline\hline
\end{tabular}

\end{table}

It was pointed out in Ref.~\cite{Alarcon:2012nr} that the virtual decuplet baryons may increase the
predicted pion-nucleon and decrease the strangeness-nucleon sigma terms. In addition, in Ref.~\cite{Ren:2012aj}, the predicted
$\sigma_{\pi N}$ is indeed smaller than that of the NNLO O+D BChPT~\cite{MartinCamalich:2010fp} while the $\sigma_{SN}$ is larger.
 It is interesting to check whether similar effects still exist at N$^3$LO.

  In Table~\ref{sigmaresults}, we tabulate the predicted $\sigma_{\pi B}$ and $\sigma_{s B}$ by
  the N$^3$LO O+D BChPT with the Fit-I, Fit-12, and Fit-13 LECs using the Feynman-Hellmann theorem. The only-octet results from Ref.~\cite{Ren:2012aj} are also listed for the sake of comparison. The inclusion of the virtual decuplet baryons only slightly increases the pion-baryon sigma terms and the strangeness-nucleon sigma term and decreases the other strangeness-baryon sigma terms.

We have checked if a smaller strangeness-nucleon sigma term, such as $21\pm6$ MeV as predicted in Ref.~\cite{Shanahan:2012wh}, were
used as a constraint, the $\chi^2/\mathrm{d.o.f}$ from a simultaneous fit of data Set-I and the small $\sigma_{SN}$ would increase to about $3$. In Ref.~\cite{Shanahan:2013cd}, using the PACS-CS data as an example, the influence of scale setting in LQCD simulations on the values of sigma terms is discussed. A comprehensive study of all the other LQCD data may be needed to understand the effects of scale setting on the results of a global study such as that performed in the present work.

\section{Summary}\label{SecIV}

 We have calculated the virtual decuplet contributions to the ground state octet baryon masses within the EOMS BChPT up to N$^3$LO.
 Finite-volume corrections are calculated self-consistently. Through a simultaneous fit of the latest $n_f=2+1$ LQCD simulations from the PACS-CS, LHPC, HSC, QCDSF-UKQCD and NPLQCD collaborations, the effects of virtual decuplet baryons on the light-quark mass and volume dependence of
  the LQCD data are systematically studied.

It is shown that the contributions of virtual decuplet baryons affect little the light-quark mass dependence of the octet baryon masses, indicating that their effects can not
 be easily disentangled from those of the virtual octet baryons and the tree-level diagrams. On the other hand, a slightly better description of FVCs
 can be achieved once the virtual decuplet baryons are taken into account, especially for the lattice data with $M_{\phi}L<4$.
 This is demonstrated by a careful study of the NPLQCD and QCDSF-UKQCD small-volume data.

 Regarding the pion- and strangeness-baryon sigma terms of the octet baryons, it is shown that
 at N$^3$LO, the effects of the virtual decuplet baryons are small. Effects of lattice scale setting may
 need to be studied to understand the relatively large $\sigma_{SN}$ predicted in the present work.

\begin{acknowledgments}
XLR and LSG acknowledge useful discussions with Hua-Xing Chen and Jorge Martin Camalich.
This work was partly supported by the National
Natural Science Foundation of China under Grants No. 11005007, No. 11035007, and No. 11175002,  the New Century Excellent Talents in University  Program of Ministry of Education of China under Grant No. NCET-10-0029,  and
the Research Fund  for the Doctoral Program of Higher Education under Grant No. 20110001110087.

\end{acknowledgments}

\bibliographystyle{apsrev4-1}
\bibliography{decuplet-EOMS}

\end{document}